\documentstyle[twoside,fleqn,psfig,hyperon99_paper]{article}

\newcommand{\AmS}{{\protect\the\textfont2
  
A\kern-.1667em\lower.5ex\hbox{M}\kern-.125emS}}

\title{Polarization of Inclusive $\Lambda_c$'s in a Hybrid Model}

\author{Gary R. Goldstein
\address{Department of Physics,
        Tufts University\\
Medford, MA 02155}%
        \thanks{This work is supported in part by funds
provided by the U.S. Department of Energy (D.O.E.)
\#DE-FG02-92ER40702.}}
       
\begin{document}

\begin{abstract}
A hybrid model is presented for hyperon polarization that is based on
perturbative QCD subprocesses and the recombination of polarized quarks
with scalar diquarks. The updated hybrid model is applied to
$p+p\rightarrow \Lambda +X$ and successfully reproduces the detailed
kinematic dependence shown by the data. The hybrid model is extended
to include pion beams and polarized $\Lambda_c$'s. The resulting
polarization is found to be in fair agreement with recent experiments.
Predictions for the polarization dependence on $x_F$ and $p_T$ is given. 
\end{abstract}

% typeset front matter (including abstract)
\maketitle

\section{INTRODUCTION}
%\documentstyle[10pt]{article}    
%simple latex w/o style files

\author{Gary R. Goldstein
\thanks{This work is supported in part by funds  
provided by the U.S. Department of Energy (D.O.E.) 
\#DE-FG02-92ER40702.}\\
Department of Physics\\
Tufts University\\
Medford, MA 02155  USA}

Inclusively produced strange hyperons can have sizeable  
polarization~\cite{heller1} over a wide range of energies. Evidence now
indicates that charmed hyperons also have sizeable
polarization~\cite{accmor,e791}. Many theoretical models have
been proposed to explain various aspects of hyperon polarization
data with
varying success~\cite{models,degrand,lund}. All try to explain the
large negative $\Lambda$ polarization. Because the hyperon
data is in the region of high CM energy but relatively small
transverse momentum ($p_T\sim 1$ GeV/c), soft QCD effects should play a major
role in any theoretical explanation. Several years ago Dharmaratna
and Goldstein developed a hybrid model for $\Lambda$ 
polarization in inclusive reactions~\cite{dharma1}. The model involves
hard scattering at the parton level, gluon fusion and light quark pair  
annihilation, to produce a polarized heavy quark which
then undergoes a soft recombination that, in turn, enhances the
polarization of the hyperon. This scheme provided an explanation for the
characteristic kinematic dependences of the polarization in
$p + p \rightarrow \Lambda + X$. The use of perturbative QCD to produce
the initial polarization for strange quarks, with their low current
or constituent quark mass (compared to $\Lambda_{QCD}$) made the
application of perturbation theory marginal, however. 

In the heavy quark realm the perturbative
contribution is more reliable. Given these circumstances, I have modified
the original hybrid model to apply to heavy flavor baryons produced inclusively
from either proton or pion beams. The results are encouraging, as the
following will show (see ref.~\cite{grg} for a more complete treatment).

\section{HYBRID MODEL}

%\vspace{5mm}

\begin{figure}[hctb]
\centerline{\psfig{figure=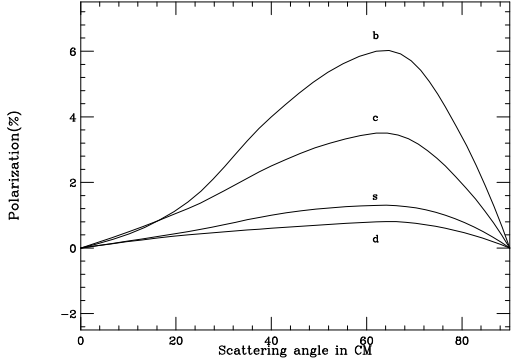,width=6.7cm}}

\caption{Polarization for the QCD subprocess of gluon fusion to quark
pairs. The curves are for d, s, c, b quarks.}
\label{fig:gg11}
%\label{fig:largenenough}
\end{figure}

All of the models for $\Lambda$ polarization begin with the observation
that $Q$-flavor hyperons of the type $\Lambda_Q \sim [ud] Q$ have their  
polarization carried primarily by the $Q$; the  
[ud] must be a color anti-triplet isospin 0 spin scalar
diquark (to the extent that $gluons + {\bf L} + sea$ contributions can be
ignored). How does the $Q$ itself get polarized in a production process?
Consider $parton + parton \rightarrow Q_{\uparrow} + \bar{Q}$.
At tree level in QCD, there can be no single quark polarization for these 
two-body subprocesses, all
diagrams being relatively real.  This can be seen when the polarization is
written in terms of helicity amplitudes $f_{a,b,c,d}$ for particles  
$A + B \rightarrow C + D$ as
\begin{eqnarray} 
{\mathcal P}_Q & \propto & \sum_{a,b,d} f_{a,b;c,d}^*
f_{a,b;c',d}
({\bf \sigma \cdot \hat{n}})_{c,c'} \nonumber \\
               & \propto & Im \sum f_{a,b;+,d}f_{a,b;-,d}^*,
\label{density}
\end{eqnarray}
where $\bf{\hat{n}}$ is the normal to the scattering plane.
Hence there has to be a phase difference and a flip--non-flip  
interference. In QCD with zero quark masses there are only  
non-flip vertices; helicity flip requires non-zero quark masses. And a  
relative imaginary part only arises beyond tree level~\cite{kane}. So 
the hybrid model incorporates the order $\alpha_s^2$ QCD
perturbative calculation of interference between tree level and the
large number of one loop diagrams to produce massive heavy quark
polarization. (Only the imaginary parts of the one loop diagrams were
needed, so the Cutkosky rules were used to
simplifiy the calculation. For the lengthy results see
ref.~\cite{dharma2,dharma3} as well as an independent calculation in
ref.~\cite{brandenburg}.) This gives rise to
significant polarization~\cite{dharma3}, proportional to 
$\alpha_s(Q^2)$ and to complicated functions of the constituent quark
mass. The scale here is $Q^2 \sim m_Q^2 \gg \lambda_{QCD}^2$.
The results are illustrated in fig.~\ref{fig:gg11} for the $g + g
\rightarrow Q_{\uparrow} +
\bar{Q}$ case, with CM energy 26 GeV and outgoing quark flavors $Q =
d,s,c,b$. The symmetry requires ${\mathcal P}(\pi-\theta) = -
{\mathcal P}(\theta)$, so backward
$Q$ has ${\mathcal P} < 0$. The magnitude of ${\mathcal P}$ reaches $\sim
6\%$ for the b-quark.
It is clear that the polarization increases roughly as the quark
mass. Similar results are obtained for $q + \bar{q} \rightarrow
Q_{\uparrow} + \bar{Q}$.
%%%%%% figure starts

The cross sections for polarized $Q$-quarks (polarized normal to the
production plane) must then be convoluted with the relevent structure
functions for the hadronic beam and target. The inclusive cross section
for $hadron + hadron \rightarrow Q(\uparrow {\mathrm or} \downarrow) + X$
is obtained thereby. For protons on protons gluon fusion is the more
significant subprocess.

The hadronization process, by which the polarized $Q$ recombines with
a [ud] diquark system to form a $\Lambda_Q$, is
crucial for understanding the subsequent hadron polarization. The
backward moving, negatively polarized heavy quark must be accelerated to
recombine
with a fast moving diquark (resulting from remnants of the $p p$ or $\pi
p$ collision) to form the hadron with particular $x_F$ while preserving
the quark's $p_T$ value. Letting
$x_Q$ be the Feynman $x$ for the heavy quark, the simple form, a linear
mapping of the $Q$ kinematic region,
\begin{equation}
x_F = a + bx_Q
\label{xfeqn}
\end{equation}
is used for the recombination. Naively, if the Q has 1/3 of the
final hyperon momentum (in its infinite momentum frame) and the diquark
carries 2/3 of that momentum,
then $a=2/3$ and $b=1$. The values actually used, $a=0.86$ and $b=0.70$,
were chosen to fit the $pp\rightarrow \Lambda+X$ data (that
existed in 1990) at one $x_F$ value. These parameters in eqn.~\ref{xfeqn}
are not far from the naive expectation. 

This recombination
prescription is similar to the semi-classical dynamical mechanism used in the 
``Thomas precession'' model of hyperon polarization~\cite{degrand},
which posits that the s-quark needs to be accelerated by a confining
potential or via a ``flux tube''~\cite{lund} at an angle 
to its initial momentum in order to join with the diquark to form the
hyperon. The skewed acceleration gives rise to a spin precession for the
s-quark. With the precession rate, ${\bf \omega_T} = (\gamma-1){\bf v
\times
a}/v^2 \propto {\bf p_Q \times \Delta p_L}  \sim -\bf{\hat{n}}$, an energy
shift $-{\bf S\cdot \omega_T} \propto + {\bf S\cdot\hat{n}}$ occurs. Hence 
negative values of $\langle{\bf S\cdot\hat{n}}\rangle$ are energetically
favored. In the Hybrid Model the $Q$ has acquired
negative polarization already from the hard subprocess before it is
accelerated in the hadronic recombination process. That
``seed'' polarization gets
enhanced by a multiplicative factor $A\simeq 2\pi$ which simulates the
Thomas precession. The Hybrid Model combines hard
perturbative QCD with this simple model for non-perturbative
recombination. 

In summary, the hyperon polarization is given as
\begin{equation}
{\mathcal P}_{\Lambda_Q}(x_F,p_T) = A \cdot {\mathcal P}_Q(x_Q(x_F),p_T)
\label{peqn}
\end{equation}
for each reaction $g(x_1)+g(x_2)$ or $q(x_1)+\bar{q}(x_2) \rightarrow
Q\bar{Q}$, with the
mapping function $x_Q(x_F)$ obtained by inverting eqn.~\ref{xfeqn}. 
From eqn.~\ref{peqn} the subprocess polarized cross sections,
\begin{equation}
\frac{d^2\sigma(\uparrow {\mathrm and} \downarrow)_{i,j}}{dx_Qdp_T}
\nonumber
\end{equation}
for partons $(i,j)$ at $(x_1,x_2)$ can be obtained.
These cross sections are convoluted with the gluon,
quark and antiquark structure functions for
the proton and pion~\cite{duke}, $g^{p,\pi}(x), q^{p,\pi}(x),
\bar{q}^{p,\pi}(x)$, or generically $f_i^{p,\pi}(x)$ leading to
\begin{eqnarray}
\frac{d^2\sigma(\uparrow {\mathrm and} \downarrow)}{dx_Qdp_T} & =
& \sum_{i,j}\int_0^1 dx_1\int_0^1 dx_2 f_i^{p,\pi}(x_1)
\nonumber \\
 & & \cdot f_j^p(x_2)\frac{d^2\sigma(\uparrow {\mathrm and}  
\downarrow)_{i,j}}{dx_Qdp_T}.
\label{dsigma}
\end{eqnarray}
Next the recombination formula, eqn.~\ref{xfeqn}, is applied to obtain
the corresponding $\Lambda_Q$ polarized cross section at $x_F(=a+bx_Q)$
and $p_T$. The polarization is obtained via
\begin{equation}
{\mathcal P}_{\Lambda_Q}(x_F,p_T) =
A\frac{d^2\sigma(\uparrow) - d^2\sigma(\downarrow)}{d^2\sigma(\uparrow)
+ d^2\sigma(\downarrow)},
\label{polzn}
\end{equation}
in an obvious notation.

Note that the linear form of eqn.~\ref{xfeqn} maps the $Q$-quark Feynman x
region
$[-1,(1-a)/b]$ into the $x_F$ region $[(a-b),+1]$ for the $\Lambda_Q$. The
$p+p\rightarrow Q$
differential cross section, $d^2\sigma/dx_Qdp_T$ is mapped
correspondingly into the $p+p\rightarrow \Lambda_Q$ cross section
$d^2\sigma/dx_Fdp_T$. The measured cross sections for the latter are
known to fall with positive $x_F$ and to fall
precipitously with $p_T$, roughly as
\begin{equation}
(1-x_F)^{\alpha}e^{-\beta p_T^2}
\label{cross}
\end{equation}
overall~\cite{accmor}, where $\alpha$ and $\beta$ are greater than 1.0
(for $\pi+p\rightarrow\Lambda+X$ $\alpha,\beta\approx3.0$). However, the
directly computed lowest order p+p$\rightarrow$s-quark cross section grows
with $x_Q$ in the region (-1,0) and it falls more gradually with $p_T$
than the exponential in eqn.~\ref{cross}. Hence the more complete
recombination scheme would have to temper the $x_F$ dependence and
narrow the $p_T$ distribution. This will not affect the polarization
calculation, though, since the individual up or down polarized cross
sections will be alterred in the same way. For a more thorough calculation
this should be done, and work is underway on this point. The polarization
results are the focus of this work.

\section{COMPARISON WITH DATA AND PREDICTIONS}

\begin{figure}[hctb]
\centerline{\psfig{figure=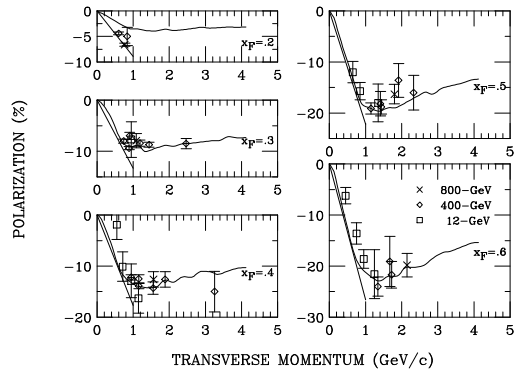,width=6.7cm}}

\caption{Hybrid model $\Lambda$ polarization in $p+p\rightarrow \Lambda +
X$ as a
function of $p_T$ for various values of $x_F$. The data at 12
GeV~\cite{heller3}, 400 GeV~\cite{lundberg}, 800 GeV~\cite{ramberg}
are shown. Exclusive data at 27.5 GeV/c~\cite{felix} is approximated by
the straight line from the origin to $p_T\approx 1$ GeV/c.}
\label{fig:gg10}
\end{figure}

Applied to strange $\Lambda$ production, the hybrid model reproduces the
detailed $(x_F,p_T)$ dependence of the data, with very slow energy
dependence~\cite{heller3}, as fig.~\ref{fig:gg10} shows.

Note that an estimated 20 to 30\% of the $\Lambda$'s come from
$\Sigma^0\rightarrow\gamma \Lambda$~\cite{lundberg}, so the parameter $A$
in eqn.~\ref{polzn} is increased to 7.9. The agreement of the hybrid model
with the wide range of data is excellent.

It is worth remarking that recently extensive data have been collected on
$\Lambda$
polarization in many {\it exclusive\/} reactions~\cite{felix}, for
which a simple form, $P = (-0.443\pm 0.037)x_Fp_T$, approximates all the
polarization data at $p_{lab}=27.5$ GeV/c. That form provides lower
bracketing values for the inclusive
polarization, as fig.~\ref{fig:gg10} indicates. In the hybrid model
all the final states other than the $\Lambda$ arise from the hadronization
of the $\bar{s}$-quark
and the remains of the incoming baryons. Therefore, in the hybrid model it
would be anticipated that as the beam energy increases and/or more final
states are
included in the determination of the $\Lambda$ polarization, more
complicated final states will be accompanied by much lower polarization as
$p_T$ increases beyond 1 GeV/c.

\begin{figure}[hctb]
\centerline{\psfig{figure=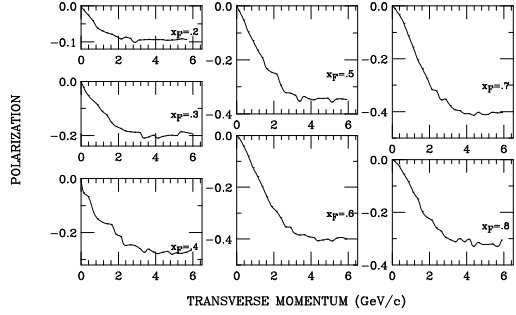,width=6.7cm}}

\caption{
%\vspace{3in}
%\special{psfile=fig13gg99.ps hscale=90 vscale=90 hoffset=-30
%voffset=-200}
%\vspace{0.5in}
%\caption{
$\Lambda_c$ polarization in $\pi^- + p\rightarrow
\Lambda_c + X$ as a function of $p_T$ for various values of $x_F$.
Multipying these polarizations by $m(\Lambda_c)/m(\Lambda)$ will
incorporate the hadron mass enhancements as in fig.~\ref{fig:gg12}. }
\label{fig:gg13}
\end{figure}

\begin{figure}[hctb]
\centerline{\psfig{figure=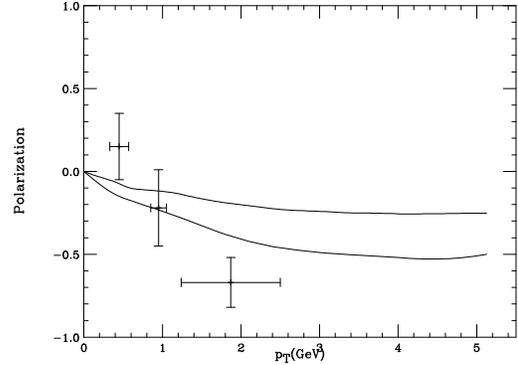,width=6.7cm}}

\caption{
%\vspace{3in}
%\special{psfile=fig12gg99.ps hscale=60 vscale=60 voffset=-100}
Estimate of $\Lambda_c$ polarization from $\pi^{-} p\rightarrow
\Lambda_c + X$. The larger polarization includes heavy mass enhancements.
The preliminary data~\cite{e791} is from E791.}
\label{fig:gg12}
\end{figure}

In turning to $\Lambda_c$ production, there is a straightforward scaling
up that occurs in the $\mathcal{P}$ equations for $g+g$ and
$q+\bar{q}\rightarrow c\uparrow +\bar{c}$. The seed polarization increases
by $\sim 3$. The recombination with a fixed force/mass should have the
same Thomas factor, but the overall recombination could scale as
$M_{hadron}$, so a factor of $M_{\Lambda_c}/M_{\Lambda} \sim 2$ could
apply. The scaled polarization in the reaction $\pi + p \rightarrow
\Lambda_c\uparrow + X$ is obtained from the convolution of
eqn.~\ref{dsigma} with the $\pi$ structure functions for the
beam~\cite{duke}. The predicted kinematic dependences for
${\mathcal P}(x_F,p_T)$ are shown in fig.~\ref{fig:gg13} (without the
hadron mass enhancement). 
Integrating over
$x_F$ from -0.2 to +0.6 allows the comparison with the data of
E791~\cite{e791,george}, as fig.~\ref{fig:gg12} shows. The lower curve has
taken
the additional factor of 2 that could apply to the scaling of the
recombination. The higher curve does not have that factor and gives a
poorer fit, albeit not far from the large uncertainties in the data
points.
 \section{CONCLUSION}
In conclusion, these results are encouraging for the hybrid model.
The Thomas enhanced gluon fusion  
model has been modified to include quark-anti-quark annihilation, which
should be more prominent for heavy baryon polarization in pion induced
reactions, like the above $\pi^- + p \rightarrow \Lambda_c + X$.
Experimental data can be analyzed into $x_F$ as well as $p_T$ bins, so the
predictions from the hybrid model can be checked in detail. It is
important to realize that the results for the $\Lambda_c$ were obtained
without changing the parameters of the model that had been applied to the
strange hyperons. Aside from the possible enhancement in $A$, everything
else was simply scaled up by quark mass. This gives further credence to
the results herein.

The somewhat
{\it ad hoc\/} prescription for the recombination is being studied further
in order to accommodate both the polarization and the cross section
behavior of eqn.~\ref{cross}, with the kinematic variables $x_F$ and
$p_T$. The overall factor $A$ may have some dependence on those variables
as well, given that the semi-classical Thomas precession may have such
dependence. Furthermore, an investigation of other hyperon production
reactions, involving $\Sigma, \Sigma_c$, and $\Xi$, for example, is
underway. Will $\bar{p} + p \rightarrow \Lambda + X$ carry
significant, near energy independent polarization at collider energies?
Can photoproduction of $\Lambda$ produce large polarizations also? These
can be answered within the hybrid model. 

The related strange meson asymmetries in $p\uparrow + p \rightarrow K
\ {\mathrm or}\ \pi \ {\mathrm or}\ \Lambda $ will be investigated in
future work as well. 

\section{ACKNOWLEDGMENTS}
The author thanks Austin Napier and  appreciates correspondence
with members of E791, particularly M.V. Purohit, G.F. Fox and J.A. Appel. 
He is grateful to the organizers of Hyperon99 for inviting him and for
producing a lively conference.


\begin{thebibliography}{9}

\bibitem{heller1} K. Heller, {\it ``Inclusive hyperon polarization: a
review''\/} {\it in\/}  Proceedings of the 6th International Symposium on
High Energy Spin Physics, Marseille, edited by J. Soffer, Les Editions de
Physique (1985), p.C2-121.

\bibitem{accmor}  S. Barlag, {\it et al.\/} (ACCMOR Collaboration),
Phys.Lett. {\bf B325}, 531 (1994).

\bibitem{e791} E.M. Aitala, {\it et al.\/} (E791 Collaboration) {\it
``Multidimensional Resonance Analysis of $\Lambda_c^+ \rightarrow p K^-
\pi^+$''\/}, preprint, Fermilab 1999; M.V. Purohit, contribution to this
symposium. 

\bibitem{models} J. Szwed, Phys. Lett. {\bf 105B}, 403 (1981); S.M.
Troshin and N.E. Tyurin, Sov. J. Nucl. Phys. {\bf 38}, 693 (1983); {\it
ibid}, Phys. Rev. {\bf D55}, 1265 (1997); J. Soffer and N.E.
T\"{o}rnqvist, Phys. Rev. Lett. {\bf 68}, 907 (1992).

\bibitem{degrand} T.A. De Grand and H.I. Miettinen, Phys. Rev. {\bf D24}
2419 (1981). 

\bibitem{lund}  B. Andersson, G. Gustafson, and G. Ingelman, Phys. Lett.
{\bf B85} 417 (1979).

\bibitem{dharma1} W.G.D.Dharmaratna and Gary R. Goldstein, Phys. Rev.
{\bf D41} 1731 (1990).

\bibitem{grg} Gary R. Goldstein, preprint hep-ph/9907573 (1999).

\bibitem{kane} G.L. Kane, J. Pumplin and W. Repko, Phys. Rev. Lett. {\bf
41} 1989 (1978).

\bibitem{dharma2} W.G.D. Dharmaratna, {\it ``Massive Quark Polarization in
Quantum Chromodynamics Subprocesses''\/}, Ph.D. dissertation, Tufts
University (1990).

\bibitem{dharma3} W.G.D.Dharmaratna and Gary R. Goldstein, Phys.Rev.
{\bf D53} 1073 (1996).

\bibitem{brandenburg} W. Bernreuther, A. Brandenburg and P. Uwer, Phys.
Lett. {\bf B368}, 153 (1996).

\bibitem{duke} D.W. Duke and J.F. Owens, Phys. Rev. {\bf D30}, 49 (1984).

\bibitem{heller3} K. Heller, {\it et al.\/}, Phys. Rev. Lett. {\bf 41} 607
(1978).

\bibitem{lundberg}  B. Lundberg, {\it et al.\/}, Phys.Rev. {\bf D40} 3557
(1989).

\bibitem{ramberg} E.J. Ramberg, {\it et al.\/}, Phys. Lett. {\bf B338} 404
(1994).

\bibitem{felix} J. F\'{e}lix, {\it et al.\/}, Phys. Rev. Lett. {\bf 82}
5213 (1999).

%\bibitem{heller2} K. Heller, {\it ``Spin and high energy hyperon
%production, results and prospects \/} {\it in\/} Proceedings
%of the 12th International Symposium on High Energy Spin Physics, 
%Amsterdam, edited by
%C.W. deJager {\it et al.}, World Scientific, Singapore, 1997, p.23.

%\bibitem{morelos} A. Morelos, {\it et al.\/} (E761 Collaboration), Phys.
%Rev. Lett. {\bf 71}, 2172 (1993).

\bibitem{george} G.F. Fox, private communication.

%\bibitem{mass} A.V. Efremov and O.V. Teryaev, Yad. Fiz. {\bf 36}, 242
%(1982); Sov. Journ. Nucl. Phys. {\bf 36}, 140 (1982); P.G. Ratcliffe,
%Nucl. Phys. {\bf B264}, 493 (1986).

\end{thebibliography}
\end{document}